# Fast Quantum Search Algorithm Modelling on Conventional Computers: Information Analysis of Termination Problem

Ulyanov Sergey V.[*] and Ulyanov Viktor S.[†]

[*]Institute of System Analysis and Management, Dubna State University
[*]Meshcheryakov Laboratory of Information Technologies, Joint Institute for Nuclear Research
[†]Department of Information Technologies, Moscow State University of Geodesy and Cartography (MIIGAiK)

[*]Email: ulyanovsv46_46@mail.ru
[†]Email: ulyanovik@gmail.com

**Abstract**

The simplest technique for simulating a quantum algorithm (QA) described based on the direct matrix representation of the quantum operators. This approach is stable and precise, but it requires allocation of operator's matrices in the computer's memory. Since the size of the operators grows exponentially, this approach is useful for simulation of QAs with a relatively small number of qubits (e.g., approximately 11 qubits on a typical desktop computer). Using this approach, it is relatively simple to simulate the operation of a QA and to perform fidelity analysis. A more efficient fast QA simulation technique is based on computing all or part of the operator matrices on an as needed current computational basis. Using this technique, it is possible to avoid storing all or part of the operator matrices. In this case, the number of qubits that can be simulated (e.g., the number of input qubits, or the number of qubits in the system state register) is affected by: (i) the exponential growth in the number of operations required to calculate the result of the matrix products; and (ii) the size of the state vector that is allocated in computer memory. In one embodiment, apply this approach it is reasonable to simulate up to 19 or more qubits on typical desktop computer, and even more on a system with vector architecture. Due to particularities of the memory addressing and access processes in a typical desktop computer (such as, for example, a Pentium-based Personal Computer), when the number of qubits is relatively small, the compute-on-demand approach tends to be faster than the direct storage approach. The compute-on-demand approach benefits from a study of the quantum operators, and their structure so that the matrix elements can be computed more efficiently. Effective simulation of Grover's quantum search algorithm as example on computer with classical architecture is considered.

## 1 Introduction: Algorithmic representation of the quantum operators and fast Grover's quantum search algorithms

The study portion of the compute-on-demand approach can, for some QAs lead to a problem-oriented approach based on the QA structure and state vector behavior [1-3]. For example, in Grover's quantum search algorithm (QSA) [4], the state vector always has one of the two different values: (i) one value corresponds to the probability amplitude of the answer; and (ii) the second value corresponds to the probability amplitude of the rest of the state vector. Using this assumption, it is possible to configure the algorithm using these two different values, and to efficiently simulate Grover's QSA. In this case, the primary limit is a representation of the floating-point numbers used to simulate the actual values of the probability amplitudes. After the superposition operation, these probability amplitudes are very small $\left(\frac{1}{\sqrt{2^n}}\right)$. Thus, it is possible to simulate Grover's QSA with this approach simulating 1024 qubits or more without termination condition calculation and up to 64 qubits or more with termination condition estimation based on Shannon entropy.

Other QAs do not necessarily reduce to just two values. For those algorithms that reduce to a finite number of values, the techniques used to simplify the Grover's QSA can be used, but the maximum number of input qubits that can be simulated will tend to be smaller, because the probability amplitudes of other algorithms have relatively more complicated distributions.

Introduction of an external excitation can decrease the possible number of qubits for some algorithms. In some algorithms, the entanglement and interference operators can be bypassed (or simplified), and the output computed based only on a superposition of the initial states (and deconstructive interference of the final output patterns) representing the state of the designed schedule of control gains. For example, a particular case of Deutsch-Jozsa's and Simon algorithms can be made entanglement free by using pseudo-pure quantum states [5].

The disclosure that follows begins with a comparative analysis of the temporal complexity of several representative QAs. That analysis is followed by an introduction of the generalized approach in QA simulation and algorithmic representation of quantum operators. Subsequent portions describe the structure representation of the QAs applicable to low level programming on classical computer (PC), generalizations of the approaches and introduction of the general QA simulation tool based on fast problem-oriented QAs.

The simulation techniques are then applied to a quantum control algorithm.

The matrix-based approach can be efficiently realized for a small number of input qubits. The matrix approach is used above as a useful tool to illustrate complexity issues associated with QA simulation on classical computer.

## 2 Structure of QA gate system design

As shown in Fig. 1 (a), a QA simulation can be represented as a generalized representation of a QA as a set of sequentially-applied smaller quantum gates (see in details [1]).

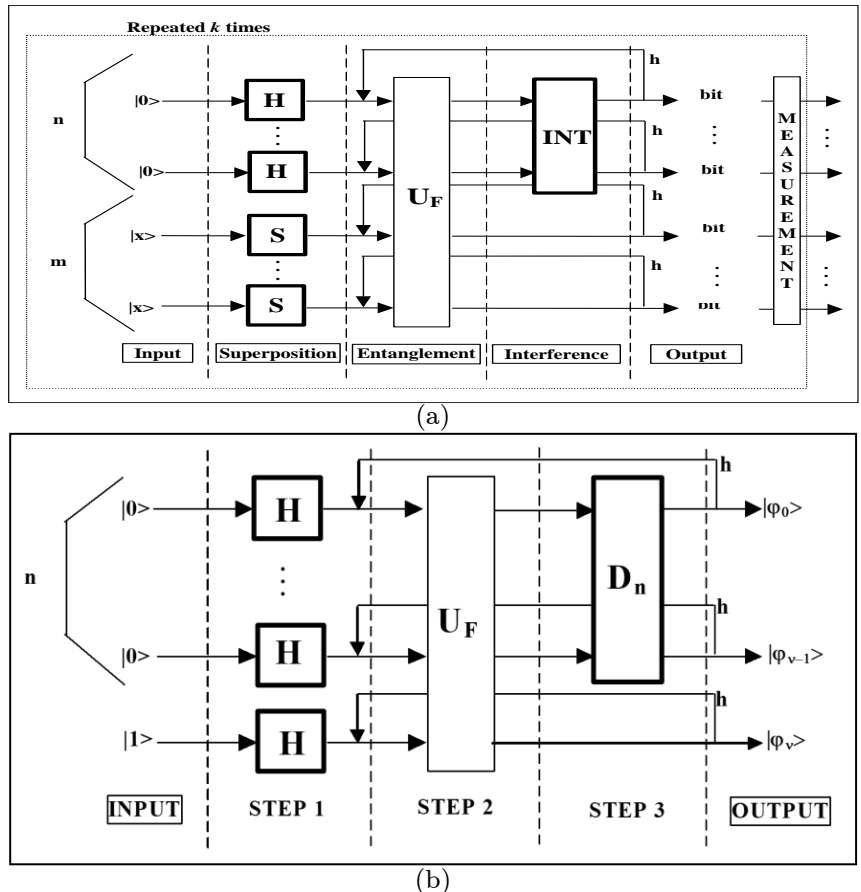

Figure 1: (a) Circuit representation of QA; (b) Quantum circuit of Grover's QSA.

Moreover, local optimization of QA components according to specific hardware realization makes it possible to develop appropriate hardware accelerators for QA simulation using classical gates.

## 3 Generalized approach in QA simulation

In general, any QA can be represented as a circuit of smaller quantum gates as shown in Figs 1 (a, b). The circuit shown in the Fig. 1 (a) is divided into five general layers: (1) input; (ii) superposition; (iii) entanglement; (iv) interference; and (v) output.

*Layer* 1: *Input.* The quantum state vector is set up to an initial value for this concrete algorithm. For example, input for Grover's QSA is a quantum state $|\phi_0\rangle$ described as a tensor product

$$
\begin{aligned}
|\phi_0\rangle = a_1|0\rangle\otimes\ldots\otimes|0\rangle\otimes|0\rangle + a_2|0\rangle\otimes\ldots\otimes|0\rangle\otimes|1\rangle + a_3|0\rangle\otimes\ldots\otimes|1\rangle\otimes|0\rangle + \ldots \\
+ a_n|1\rangle\otimes\ldots\otimes|1\rangle\otimes|1\rangle = 1|0\rangle\otimes\ldots\otimes|0\rangle\otimes|1\rangle = |0\cdots 01\rangle
\end{aligned} \quad (1)
$$

where $|0\rangle = \begin{pmatrix}1\\0\end{pmatrix}$; $|1\rangle = \begin{pmatrix}0\\1\end{pmatrix}$; $\otimes$ denotes Kronecker tensor product operation.

Such a quantum state can be presented as shown on the Fig. 2 (a).

The coefficients $a_i$ in the Eq. (1) are called probability amplitudes. Probability amplitudes can take negative and/or complex values. However, the probability amplitudes must obey the following constraint:

$$\sum_i a_i^2 = 1 \quad (2)$$

The actual probability of the arbitrary quantum state $a_i|i\rangle$ to be measured is calculated as a square of its probability amplitude value $p_i = |a_i|^2$.

*Layer* 2: *Superposition.* The state of the quantum state vector is transformed by the Walsh-Hadamard operator so that probabilities are distributed uniformly among all basis states. The result of the superposition layer of Grover's QSA is shown in Fig. 2 (b) as a probability amplitude representation, and also in Fig. 3 (b) as a probability representation.

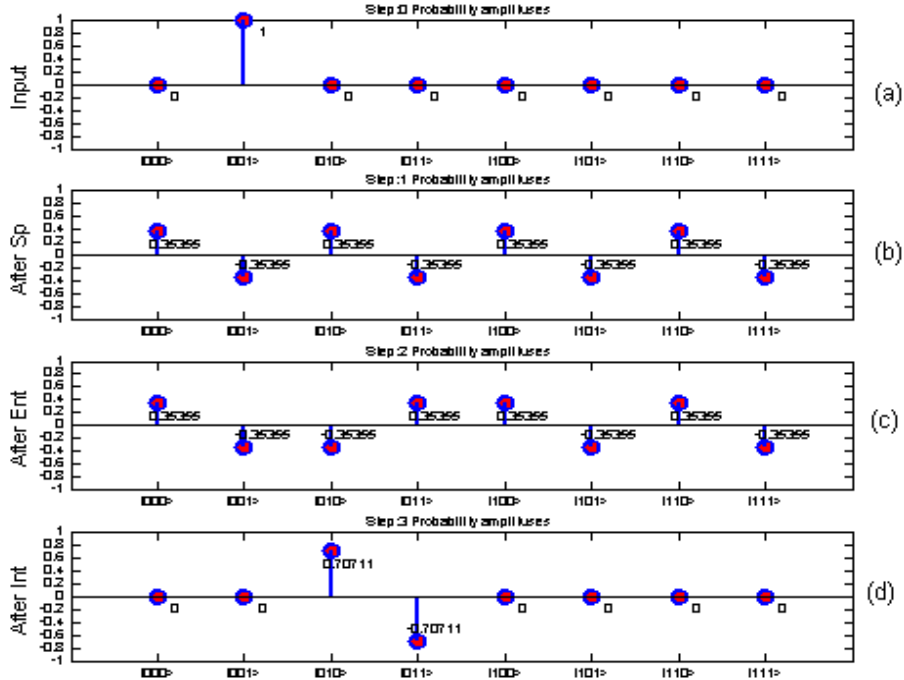

Figure 2: Dynamics of Grover's QSA probability amplitudes of state vector on each algorithm step.

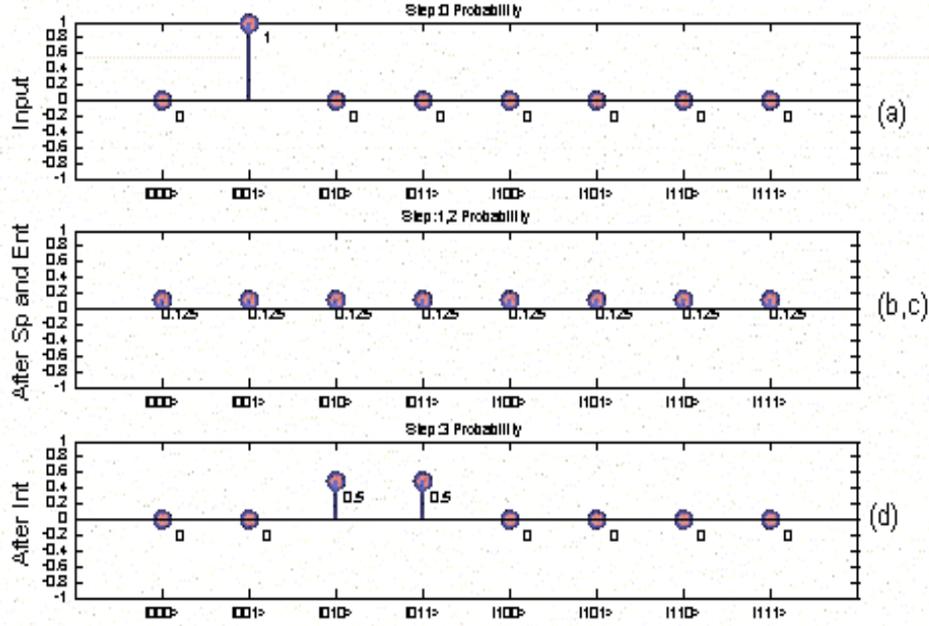

Figure 3: Dynamics of Grover's QSA probabilities of state vector on each algorithm step.

*Layer* 3: *Entanglement*. Probability amplitudes of the basis vector corresponding to the current problem are flipped while rest basis vectors left unchanged. Entanglement is typically provided by controlled-NOT (CNOT) operations. Figures 2 (c) and 3 (c) show results of entanglement from the application of the operator to the state vector after superposition operation. An entanglement operation does not affect the probability of the state vector to be measured. Rather, entanglement prepares a state, which cannot be represented as a tensor product of simpler state vectors. For example, consider state $\phi_1$ shown in the Fig. 2 (b) and state $\phi_2$ presented on the Fig. 2 (c):

$$\phi_1 = 0.35355(|000\rangle - |001\rangle + |010\rangle - |011\rangle + |100\rangle - |101\rangle + |110\rangle - |111\rangle)$$
$$= 0.35355(|00\rangle + |01\rangle + |10\rangle + |11\rangle)(|0\rangle - |1\rangle)$$,

$$\phi_2 = 0.35355(|000\rangle - |001\rangle - |010\rangle + |011\rangle + |100\rangle - |101\rangle + |110\rangle - |111\rangle)$$
$$= 0.35355(|00\rangle - |01\rangle + |10\rangle + |11\rangle)|0\rangle - 0.35355(|00\rangle + |01\rangle + |10\rangle + |11\rangle)|1\rangle.$$

As shown above, the description of state $\phi_1$ can be presented as a tensor product of simpler states, while state $\phi_2$ (in the measurement basis $\{|0\rangle, |1\rangle\}$) cannot.

*Layer* 4: *Interference*. Probability amplitudes are inverted about the average value. As a result, the probability amplitude of states «marked» by entanglement operation will increase.

Figures 2 (d) and 3 (d) show the results of interference operator application.

Figure 2 (d) shows probability amplitudes and Fig. 3 (d) shows probabilities.

*Layer* 5: *Output*. The output layer provides the measurement operation (extraction of the state with maximum probability), followed by interpretation of the result. For example, in the case of Grover's QSA, the required index is coded in the first $n$ bits of the measured basis vector.

Since the various layer of the QA are realized by unitary quantum operators, simulation of quantum operators depends on simulation of such unitary operators. Thus, in order to develop an efficient, simulation, it is useful to understand the nature of the QAs basic quantum operators.

# 4 Basic QA - operators

The superposition, entanglement and interference operators are now considered from the simulation viewpoint. In this case, the superposition operators and the interference operators have more complicated structure and differ from algorithm to algorithm. Thus, it is first useful to consider the entanglement operators, since they have a similar structure for all QAs, and differ only by the function being analyzed.

In general, the superposition operator is based on the combination of the tensor products Hadamard $H$ operators: $H = \frac{1}{\sqrt{2}}\begin{pmatrix} 1 & 1 \\ 1 & -1 \end{pmatrix}$ with identity operator $I: I = \begin{pmatrix} 1 & 0 \\ 0 & 1 \end{pmatrix}$.

*Remark.* As described in [1-3] the simulation system of quantum computation is based on quantum algorithm gates (QAG). The design process of QAG includes the matrix design form of three quantum operators: superposition *(Sp)*, entanglement ($U_F$) and interference *(Int)*. In general form, the structure of a QAG can be described as follows:

$$QAG = \left[\left(Int \otimes^n I\right) \cdot U_F\right]^{h+1} \cdot \left[{}^n H \cdot \otimes^m S\right],$$

where $I$ is the identity operator; the symbol $\otimes$ denotes the tensor product; $S$ is equal to $I$ or $H$ and dependent on the problem description. One portion of the design process in QAG is the type-choice of the entanglement problem-dependent operator $U_F$ that physically describes the qualitative properties of the function $f$.

The Hadamard Transform creates the superposition on classical states, and quantum operators such as *CNOT* create robust entangled states. The Quantum Fast Fourier Transform (QFFT) produces interference. For most QAs the superposition operator can be expressed as

$$Sp = \left(\bigotimes_{i=1}^{n} H\right) \otimes \left(\bigotimes_{i=1}^{n} S\right), \qquad (3)$$

where $n$ and $m$ are the numbers of inputs and of outputs respectively. The operator $S$ depends on the algorithm and can be either the Hadamard operator $H$ or the identity operator $I$. The numbers of outputs $m$ as well as structures of the corresponding superposition and interference operators are presented in Table 1 for different QAs.

| Algorithm | Superposition | $m$ | Interference |
|---|---|---|---|
| Deutsch's | $H \otimes I$ | 1 | $H \otimes H$ |
| Deutsch-Jozsa's | ${}^n H \otimes H$ | 1 | ${}^n H \otimes I$ |
| Grover's | ${}^n H \otimes H$ | 1 | $D_n \otimes I$ |
| Simon's | ${}^n H \otimes {}^n I$ | $n$ | ${}^n H \otimes {}^n I$ |
| Shor's | ${}^n H \otimes {}^n I$ | $n$ | $QFT_n \otimes {}^n I$ |

Table 1. Parameters of superposition and interference operators of main quantum algorithms.

## 4.1 Superposition and interference

Superposition and interference operators are often constructed as tensor powers of the Hadamard operator, which is called the *Walsh-Hadamard* operator. Elements of the Walsh-Hadamard operator can be obtained as

$$\left[ {}^{n}H \right]_{i,j} = \frac{(-1)^{i*j}}{\sqrt{2^n}} \left[ {}^{n-1}H \right] = \frac{1}{\sqrt{2^n}} \begin{pmatrix} {}^{(n-1)}H & {}^{(n-1)}H \\ {}^{(n-1)}H & -{}^{(n-1)}H \end{pmatrix}, \tag{4}$$

where $i = 0, 1$, $j = 0, 1$, **H** denotes Hadamard matrix of order 3.

The rule in Eq. (4) provides way to speed up of the classical simulation of the Walsh-Hadamard operators, because the elements of the operator can be obtained by the simple replication described in Eq. (4) from the elements of the ${}^{n-1}H$ order operator.

As an example, consider the superposition operator of Grover's algorithm, for the case $n = 2$, $m = 1$, $S = H$:

$$[Sp]^{Grover's} = {}^{2}H \otimes H = \left(\frac{1}{\sqrt{8}}\right)^{3} H = \left(\frac{1}{\sqrt{8}}\right)\begin{pmatrix} {}^{2}H & {}^{2}H \\ {}^{2}H & -{}^{2}H \end{pmatrix} = \left(\frac{1}{\sqrt{8}}\right)\begin{pmatrix} H & H & H & H \\ H & -H & H & -H \\ H & H & -H & -H \\ H & -H & -H & H \end{pmatrix} \tag{5}$$

Interference operators are calculated for each algorithm according to the parameters listed in Table 1. The interference operator is based on the interference layer of the algorithm, which is different for various algorithms, and from the measurement layer, which is the same or similar for most algorithms and includes the $m^{\text{th}}$ - tensor power of the identity operator.

The interference operator of Grover's algorithm can be written as a block matrix of the following form:

$$\left[ Int^{Grover's} \right]_{i,j} = D_n \otimes I = \left(\frac{1}{\sqrt{2^n}} - {}^{n}I\right) \otimes I = \left(-1 + \frac{1}{\sqrt{2^n}}\right) \otimes I \bigg|_{i,j},$$

$$\left(\frac{1}{\sqrt{2^n}}\right) \otimes I \bigg|_{i \neq j} = \frac{1}{\sqrt{2^n}} \begin{cases} -I, & i = j \\ I, & i \neq j \end{cases} \tag{6}$$

where $i = 0, \ldots, 2^n - 1$, $j = 0, \ldots, 2^n - 1$, $D_n$ refers to diffusion operator: $\left[ D_n \right]_{i,j} = \frac{(-1)^{1 \text{AND}(i=j)}}{\sqrt{2^n}}$.

For example, the interference operator for Grover's QSA, when $n = 2$, $m = 1$ is:

$$\left[ Int^{Grover's} \right]_{i,j} = D_2 \otimes I = \left(\frac{1}{\sqrt{2^n}} - {}^{2}I\right) \otimes I = \left(-1 + \frac{1}{2}\right) \otimes I \bigg|_{i=j}$$

$$= \begin{pmatrix} \left(-1+\frac{1}{2}\right)I & \frac{1}{2}I & \frac{1}{2}I & \frac{1}{2}I \\ \frac{1}{2}I & \left(-1+\frac{1}{2}\right)I & \frac{1}{2}I & \frac{1}{2}I \\ \frac{1}{2}I & \frac{1}{2}I & \left(-1+\frac{1}{2}\right)I & \frac{1}{2}I \\ \frac{1}{2}I & \frac{1}{2}I & \frac{1}{2}I & \left(-1+\frac{1}{2}\right)I \end{pmatrix} = \frac{1}{2}\begin{pmatrix} -I & I & I & I \\ I & -I & I & I \\ I & I & -I & I \\ I & I & I & -I \end{pmatrix} \tag{7}$$

As the number of qubits increases, the gain coefficient will become smaller. The dimension of the matrix increases according to $2^n$, but each element can be extracted using Eq. (6), without allocation of the entire operator matrix.

*Remark.* Since $D_n D_n^* = I$, $D_n$ is unitary and is therefore a possible quantum state transformation.

While the matrix $D_n$ is clearly unitary it can to have the decomposition form $D_n = -H_n R_n^1 H_n$, where $R_n^1[i,j] = 0$, if $i \neq j$, $R_n^1[1,1] = -1$, and $R_n^1[i,i] = +1$, if $1 < i \leq N$.

In concrete form the operator $D_n$ (*diffusion – inversion about average*) in Grover algorithm is decomposed as

$$D_n = \frac{1}{\sqrt{2^n}} \begin{pmatrix} 1 & 1 \\ 1 & -1 \end{pmatrix}^{\otimes n} \cdot \begin{pmatrix} -1 & 0 & \cdots & 0 \\ 0 & 1 & \cdots & 0 \\ 0 & 0 & \ddots & 0 \\ 0 & 0 & \cdots & 1 \end{pmatrix} \cdot \begin{pmatrix} 1 & 1 \\ 1 & -1 \end{pmatrix}^{\otimes n}$$

and can be accomplished with $O(n) = O(\log(n))$ quantum gates. It means that from the viewpoint of efficient computation the form as in Eq. (6) is more preferable.

### 4.2 Entanglement operator

The entanglement operator is a sparse matrix. Using sparse matrix operations, it is possible to accelerate the simulation of the entanglement. Each row or column of the entanglement operation has only one position with non-zero value. This is a result of the reversibility of the function $F$. For example, consider the entanglement operator for a binary function with two inputs and one output:

$$f: \{0,1\}^2 \to \{0,1\}^1, \text{ such that: } f(x) = 1\big|_{x=01} \, 0\big|_{x \neq 01}.$$

The reversible function $F$ in this case is: $F: \{0,1\}^3 \to \{0,1\}^3$, such that:

| $(x, y)$ | $(x, f(x) \oplus y)$ |
|---|---|
| 00, 0 | $00, 0 \oplus 0 = 0$ |
| 00, 1 | $00, 0 \oplus 1 = 1$ |
| 01, 0 | $01, 1 \oplus 0 = 1$ |
| 01, 1 | $01, 1 \oplus 1 = 0$ |
| 10, 0 | $10, 0 \oplus 0 = 0$ |
| 10, 1 | $10, 1 \oplus 0 = 1$ |
| 11, 0 | $11, 0 \oplus 0 = 0$ |
| 11, 1 | $11, 1 \oplus 0 = 1$ |

The corresponding entanglement block matrix can be written as:

$$U_F = \begin{matrix} & \begin{matrix} \langle 00| & \langle 01| & \langle 10| & \langle 11| \end{matrix} \\ \begin{matrix} |00\rangle \\ |01\rangle \\ |10\rangle \\ |11\rangle \end{matrix} & \begin{pmatrix} I & 0 & 0 & 0 \\ 0 & \boxed{C} & 0 & 0 \\ 0 & 0 & I & 0 \\ 0 & 0 & 0 & I \end{pmatrix} \end{matrix}.$$

Figure 2 (c) shows the result of the application of this operator in Grover's QSA.

# 5 Command line simulation of the QAs.

The example of the Grover's algorithm script is presented in Figs 4 -9.

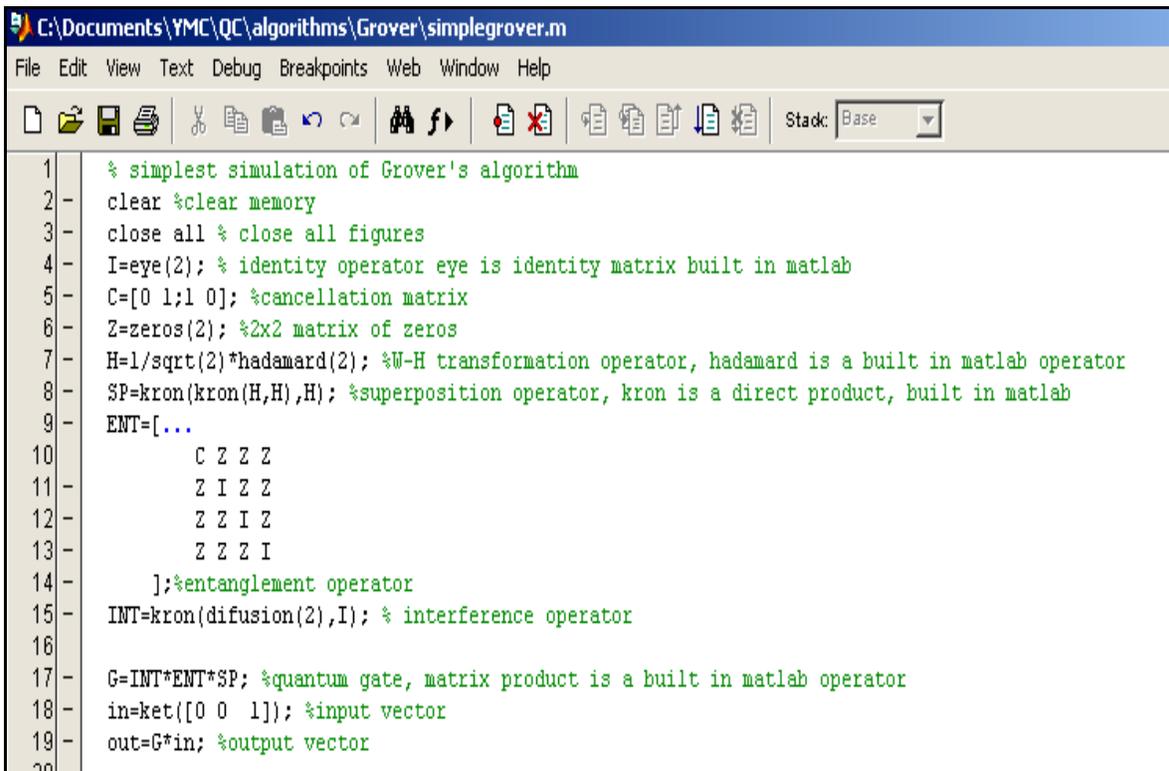

Figure 4: Example of Grover algorithm simulation script (coding of the algorithm).

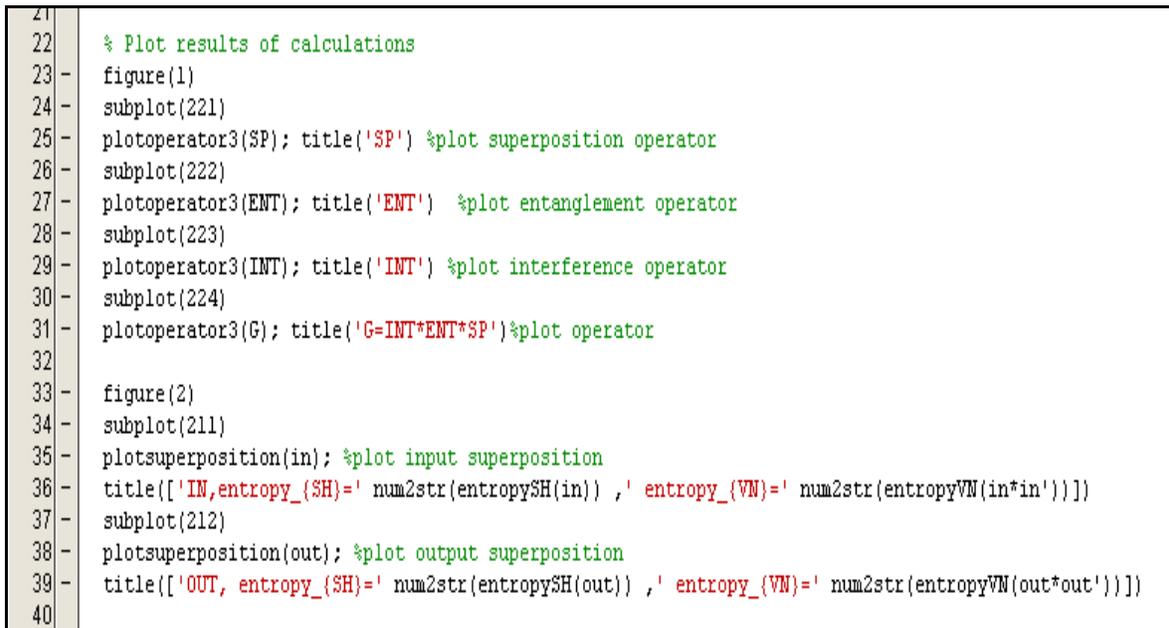

Figure 5: Example of Grover algorithm simulation script (result visualization commands).

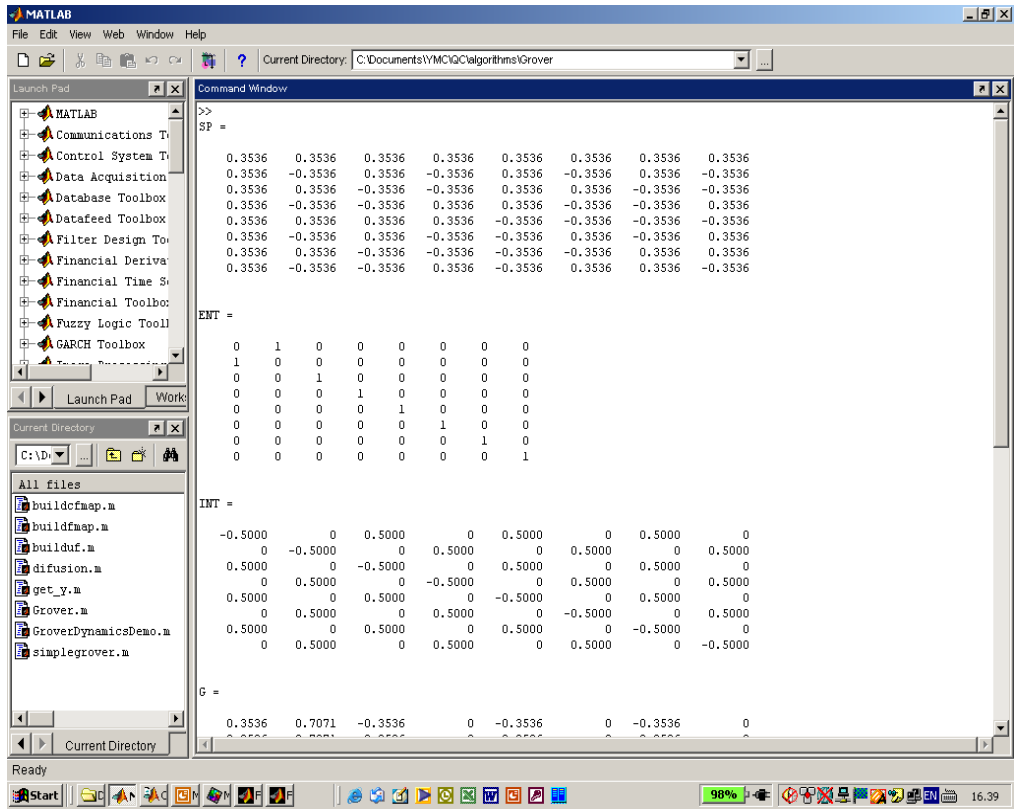

Figure 6: Example of Grover algorithm simulation script (superposition, entanglement and interference operators)

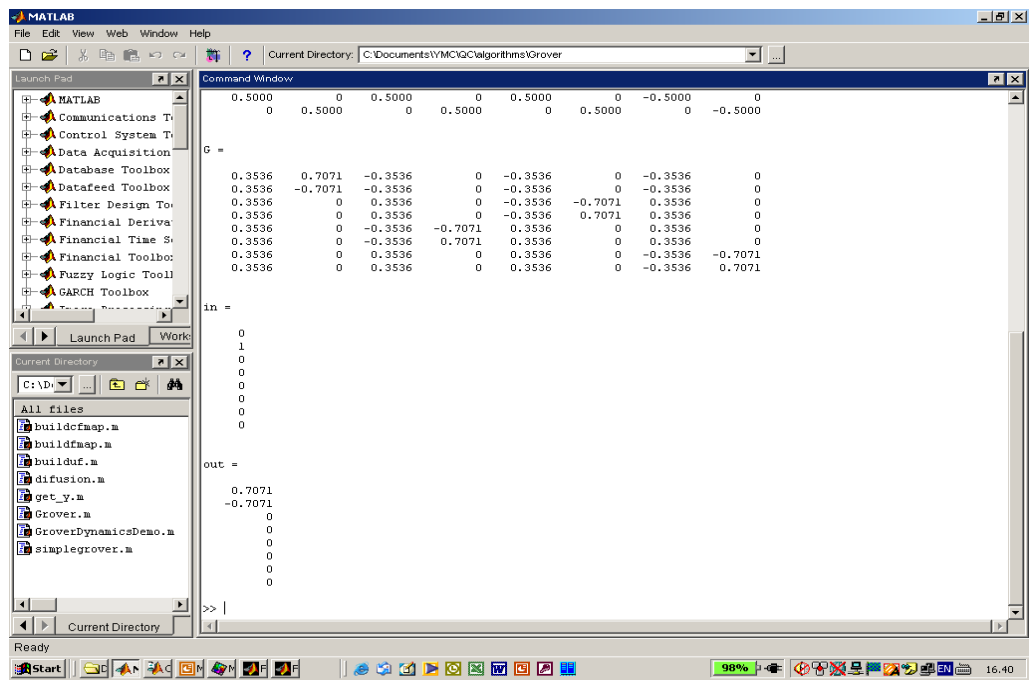

Figure 7: Example of Grover algorithm simulation script (quantum gate $G$, input vector and result of the quantum gate application).

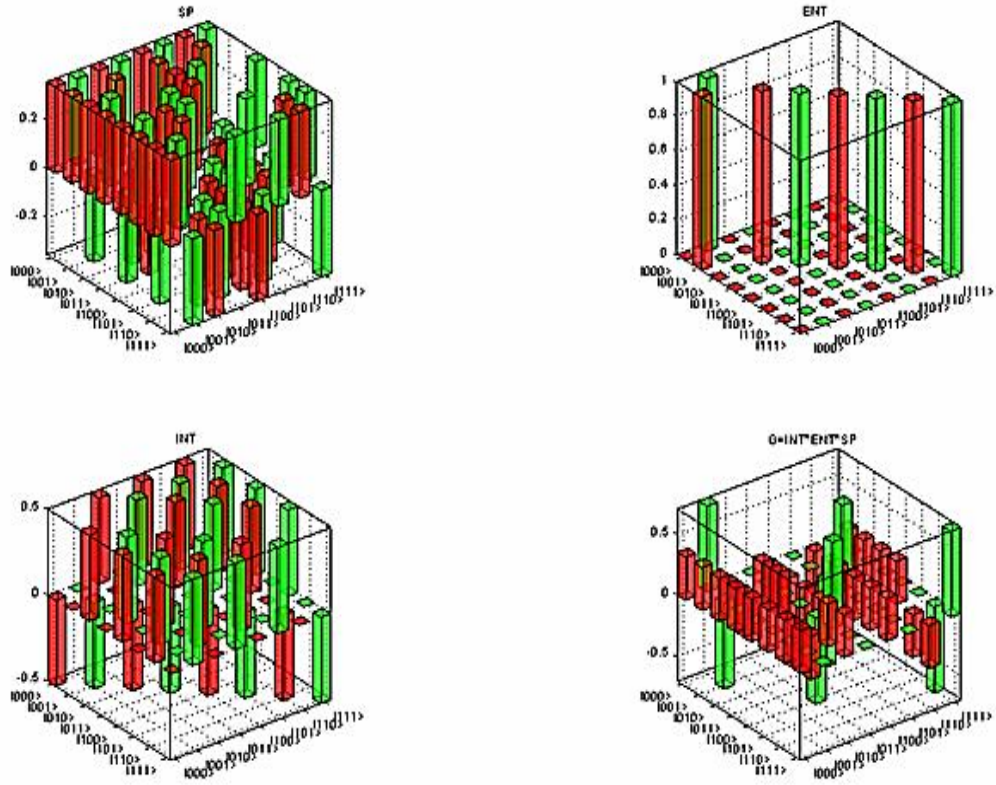

Figure 8: Example of Grover algorithm simulation script (visualization of the quantum operators)

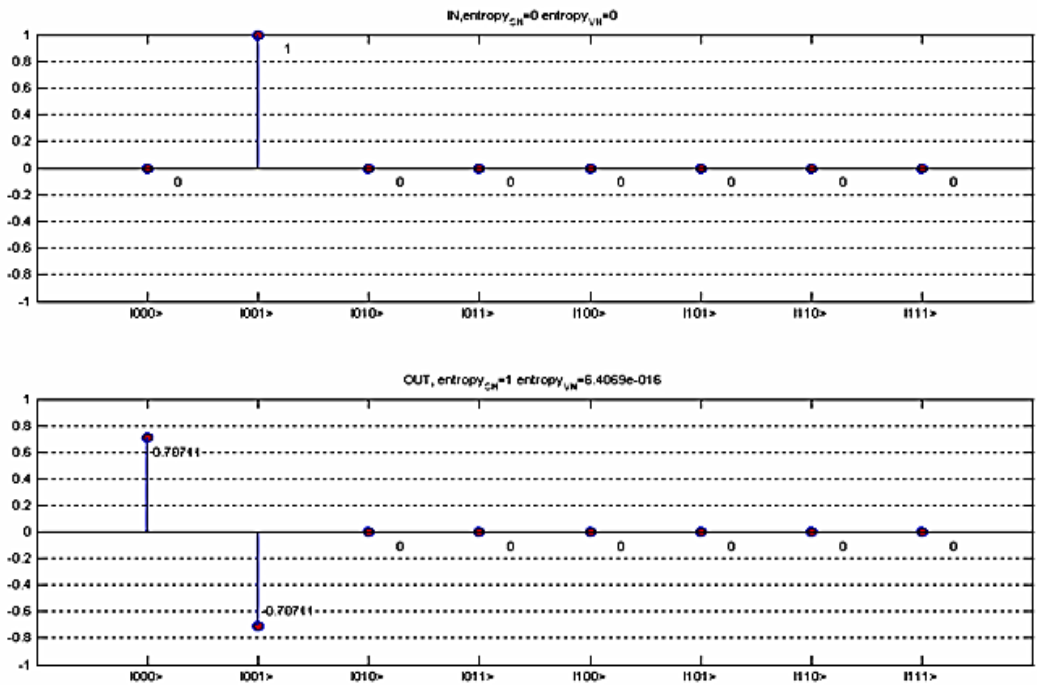

Figure 9: Example of Grover algorithm simulation script (visualization of the input and of the output quantum states)

In Fig. 4, the algorithm-related script is presented. It prepares the superposition (SP), entanglement (ENT) and interference (INT) operators of the Grover's algorithm with 3 qubits (including the measurement qubit). Then it assembles operators into the quantum gate $G$.

Then the script presented in Fig. 4 creates an input state $|in\rangle = |001\rangle$ and calculates the output state $|out\rangle = G|in\rangle$. The result of this algorithm in Matlab is an allocation of the operator matrices and of the state vectors in the memory.

Allocated quantum operator matrices are presented in Fig. 6. Allocated input $|in\rangle$ and output $|out\rangle$ state vectors as well as quantum gate $G$ are presented in Fig. 7. In order to see the results, the visualization functions are applied in Fig. 5. Code presented in Fig. 5 displays the operator matrices in Fig. 8 in 3D visualization.

In this case the vertical axis corresponds to the amplitudes of the corresponding matrix elements. Indexes of the elements are marked with the ket notation. Input $|in\rangle$ and the output $|out\rangle$ states are demonstrated in Fig. 9. In this case, the vertical axis corresponds to the probability amplitudes of the state vector components. The horizontal axis corresponds to the index of the state vector component, marked using the ket notation.

The title of the Fig. 9 contains the values of the Shannon and of the von Neumann entropies of the corresponding visualized states.

Other known QA can be formulated and executed using similar scripts, and by using the corresponding equations taken from the previous section.

# 6 Simulation of the QAs as dynamic systems

In order to simulate behavior of the dynamic systems with quantum effects, it is possible to represent the QA as a dynamic system in the form of a block diagram and then simulate its behavior in time. Figure 10 is an example of a Simulink diagram of the quantum circuit for calculation of the fidelity $\langle a|a\rangle$ of the quantum state and for the calculation of the density matrix $|a\rangle\langle a|$ of the quantum state.

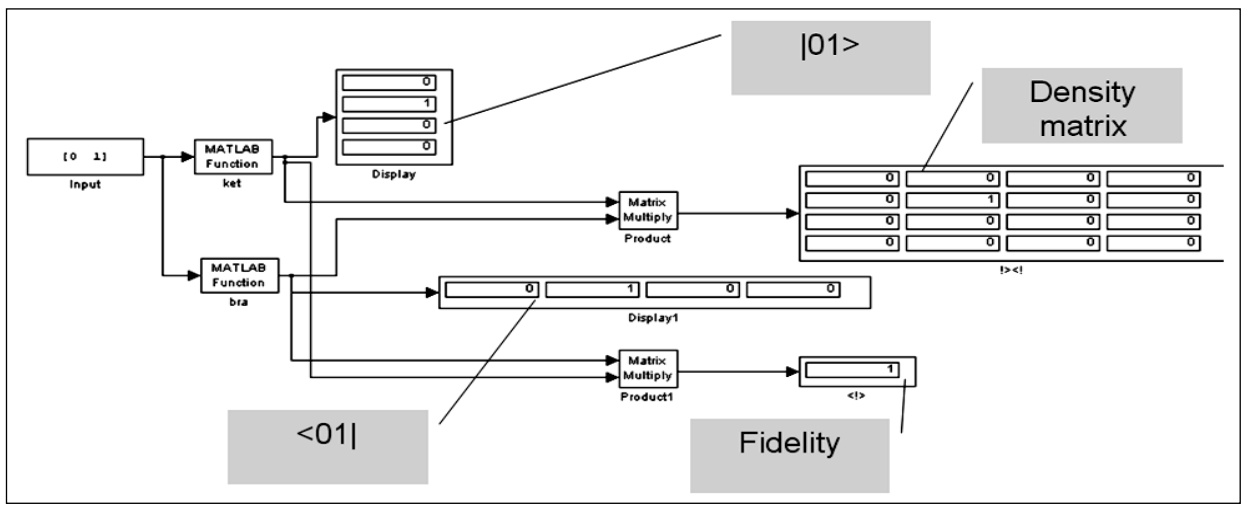

Figure 10: Simulink diagram for the simulation of the arbitrary quantum algorithm.

Bra and ket functions are taken from the common library. This example demonstrates the usage of the common functions for the simulation of the QA dynamics.

In Fig. 10, input is provided to the ket function. The output of the ket function is provided to the first input of the matrix multiplier and as a second input of the matrix multiplier. Input is also provided to the bra function. The output of the bra function is provided to the second input of the matrix multiplier and as a first input of the matrix multiplier. Output of the multiplier is a density matrix of the input state. Output of the multiplier is the fidelity of the input state.

Figure 11 shows Simulink structure of an arbitrary QA.

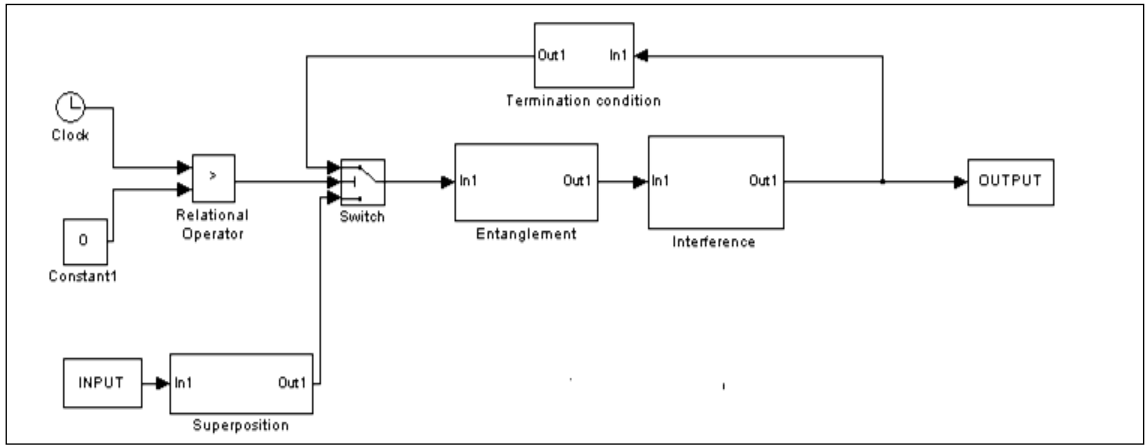

Fig. 11. Simulink diagram for the simulation of the arbitrary quantum algorithm.

Such a structure can be used to simulate a number of quantum algorithms in Matlab / Simulink environment.

## 7 Dedicated QA emulator

Developed in algorithmic representation of QAs is applicable also for design of software emulators of QAs. Key point is the reduction of the multiple matrix operations to vector operations, and following replacement of multiplication operations. This may increase dramatically emulation performance, especially on the algorithms which do not require complex number operations, and when quantum state vector has relatively simple structure (like Grover's QSA for example).

Startup window of the QA emulator is shown on the Fig. 12.

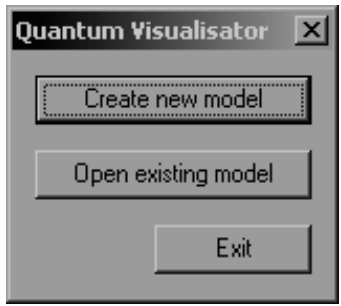 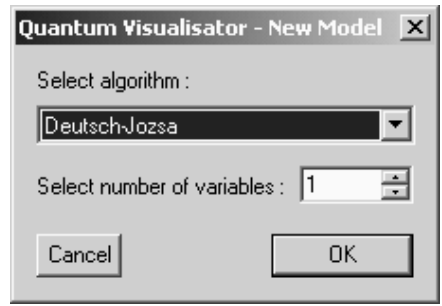

Figure 12: Start-up window of the QA emulator software.

Fig. 13. Quantum algorithm creation dialog of the QA emulator software

Here one may choose creation of the new QA model or to continue simulation of an existing one. If creation of the new model was chosen, then algorithm selection dialog (Fig. 13) will start. Here user may choose QA and its dimensions.

Actually, system may operate with up to 50 qubits and more, but due to visualization problems, it is better to limit number of qubits to 10 - 11.

Once algorithm initial parameters are set, system will draw initial state vector and selected algorithm structure in system's main window (Fig. 14).

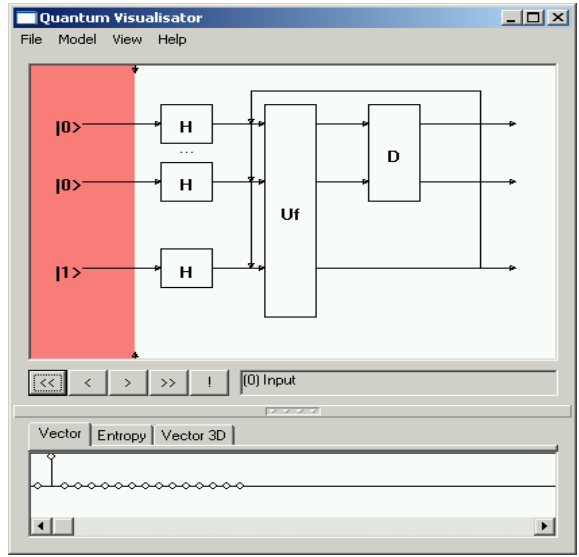

Figure 14: Main window of QA emulator software (3 qubit Grover QSA).

Main window (Fig. 14) contains all information of the emulated quantum algorithm, and permits basic operations and analysis. Form menu there is an access to involved quantum operators (Fig. 15), and it is possible to modify input functions (see below Fig. 16).

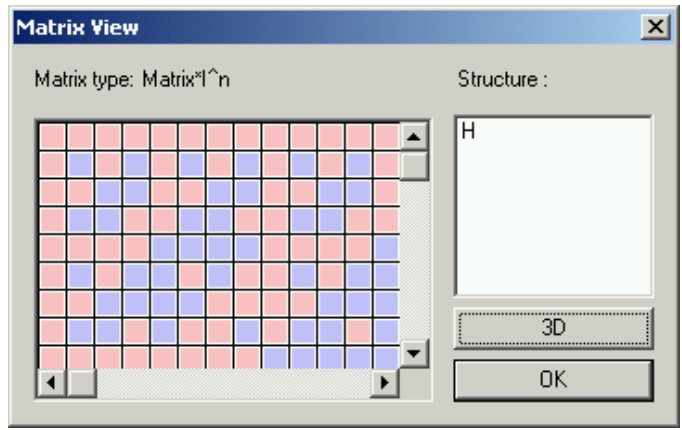

Figure 15: Plain representation of superposition operator.

QAs have reversible nature, so by clicking on arrows it is possible to make forward and backward steps of the algorithm, and currently applied algorithm step will be highlighted on the algorithm diagram.

Menu of the emulator consists of four components:

1. Item *File* provides basic operations like project save/load, and new model creation interface access;
2. Item *Model* permits an access to the input function editor (Fig. 16);
3. Item *View* provides an access to operator matrix visualizers, including Superposition, Entanglement and Interference operators. It is possible to get also 3D preview of algorithm state dynamics (Fig. 8);
4. From *Help* menu there is an access to the program documentation.

Tabbed interface in the lower part of the window permits an access to Shannon entropy chart and to 3D representation of the state vector dynamics, as well as to usual, plain representation of the QA state. Size of the tabbed area can be modified by dragging divider. Click on the middle point of divider hides tabbed area form the screen.

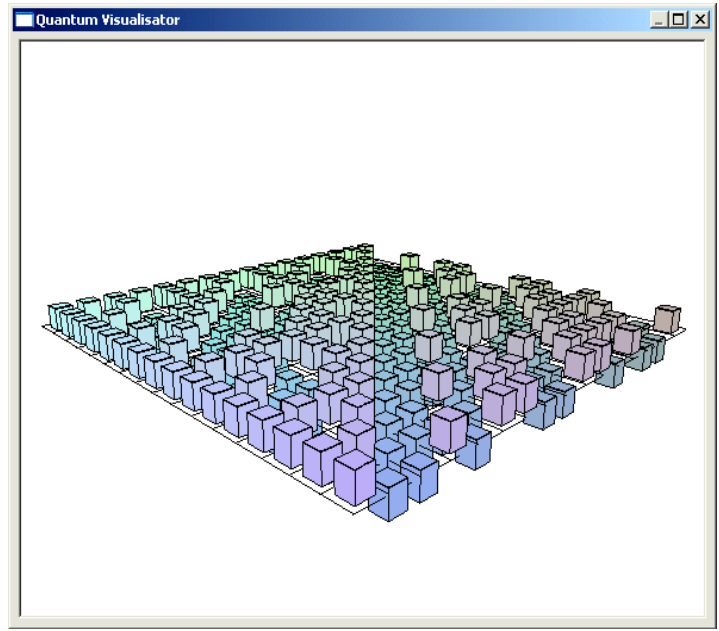

Figure 16 (a): 3D representation of superposition operator of 3-qubit Grover's QSA in quantum emulator software.

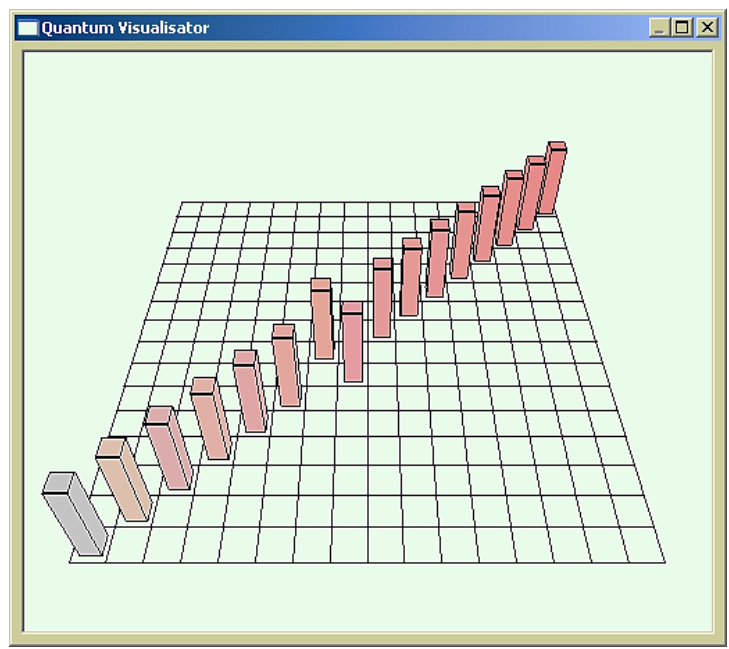

Figure 16 (b): 3D representation of entanglement operator of 3-qubit Grover's QSA in quantum emulator software.

Buttons in the middle part of the main window permit to make steps of the currently parameterized QA. As it was mentioned above, system can make forward and backward steps.

If enough steps of the algorithm were done, click on the «!¸ button will extract an answer from the current state vector.

Depending on QA an appropriate result interpretation routine will be called.

Quantum operator visualizer permits to display structure of involved quantum operator matrices in plain (Fig. 15) and in 3D (Fig. 16 (a — c)) representations.

If operator consists of tensor product of smaller operators, the possibility to have an access to sub-blocks of the tensor products is also available. 3D visuzlizer permits zoom and rotation of the charts.

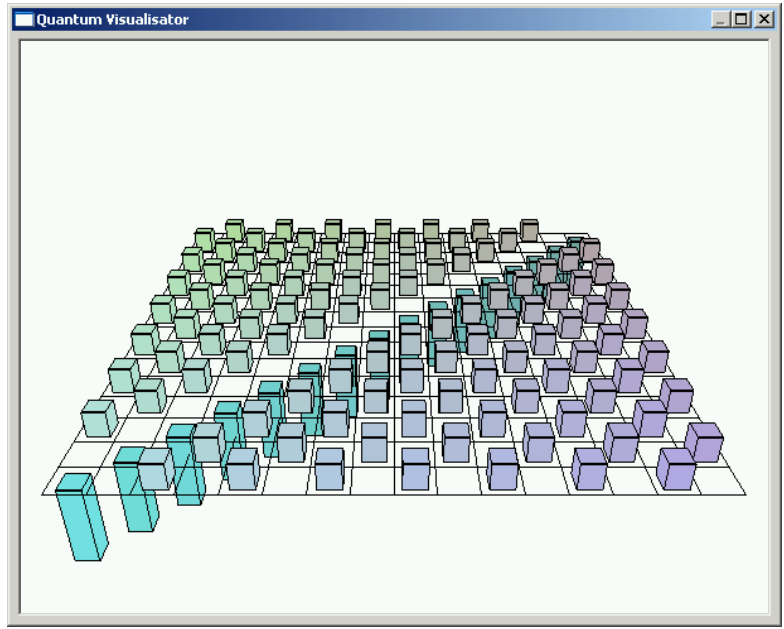

Figure 16 (c): 3D representation of interference operator of 3-qubit Grover's QSA in quantum emulator software

Input function editor permits to automate process of the entanglement operator coding as it was described in previous sections. For Grover's QSA it is possible to code functions which have more than one positive output (Fig. 17).

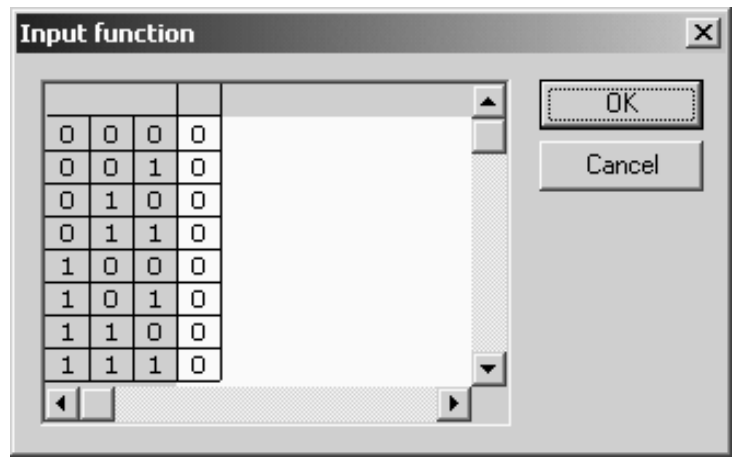

Figure 17: Input function editor of QA emulator (3-Qubit Grover's QSA).

Figures 18 - 20 show the results of Grover QSA simulation with entropy criteria termination.

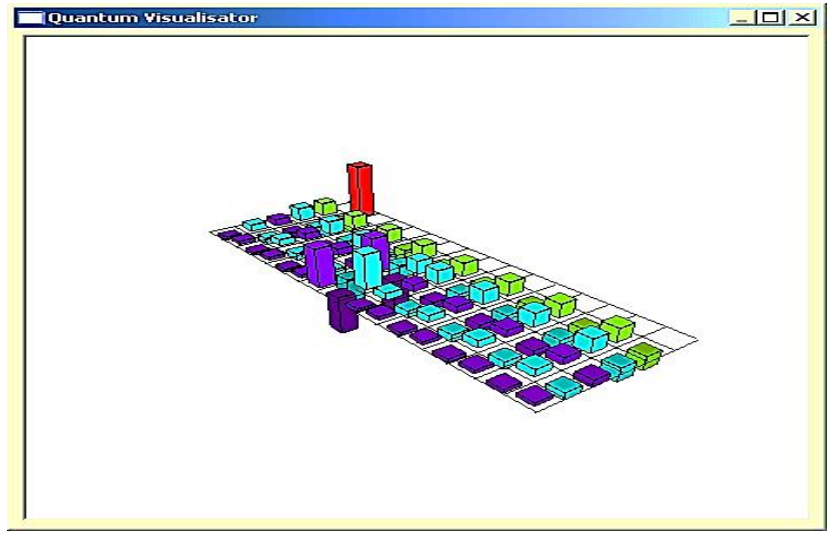

Figure 18: 3D View of 3-qubit Grover's QSA state vector after two algorithm iterations.

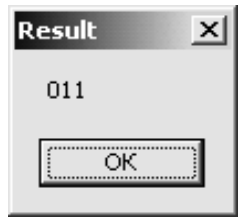

Figure 19: Answer window, case when Grover's algorithm had performed sufficient number of steps.

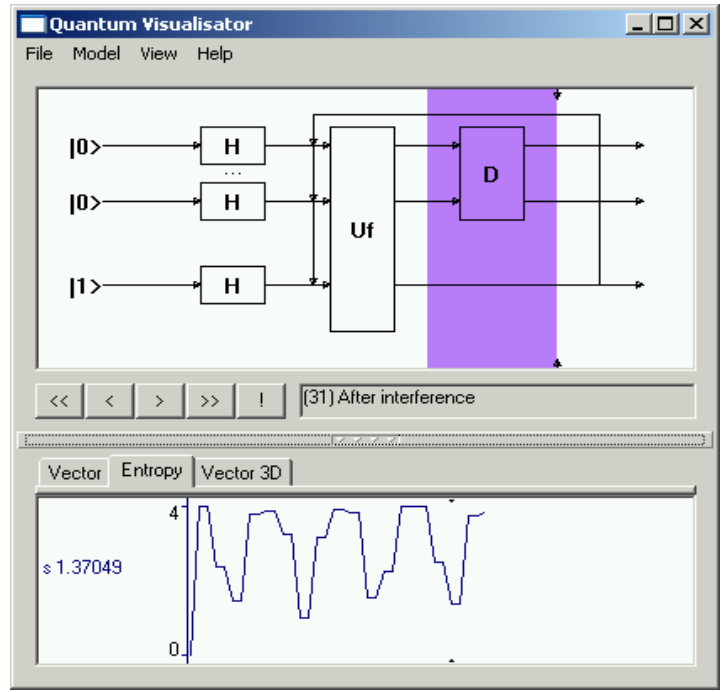

Figure 20: Shannon entropy dynamics after 31 steps of Grover's QSA/

Developed software can simulate 4 basic quantum algorithms, e.g. Deutsch-Jozsa's, Shor's, Simon's and Grover's. System uses unified easy to understand interface for all algorithms, with options of 3D visualization of state vector dynamics and of quantum operators.

Analyzing quantum operators presented in the section 5 we can do the following simplification for increasing performance of classical QA simulations: a) All quantum operators are symmetrical around main diagonal matrices; b) State vector is allocated as a sparse matrix; c) Elements of the quantum operators are not stored, but calculated when necessary using Eqs. (4), (5), (6) and (7); d) As termination condition we consider minimum of Shannon entropy of the quantum state, calculated as:

$$H^{Sh} = -\sum_{i=0}^{2^{m+n}} p_i \log p_i . \qquad (8)$$

Calculation of the Shannon entropy is applied to the quantum state after interference operation [6,7].

## 8 Results of classical quantum algorithmic gate simulation

Minimum of Shannon entropy Eq. (8) corresponds to the state when there are few state vectors with high probability (states with minimum uncertainty). Selection of appropriate termination condition is important since QAs are periodical. Figure 21 shows results of the Shannon information entropy calculation for the Grover's algorithm with 5 inputs.

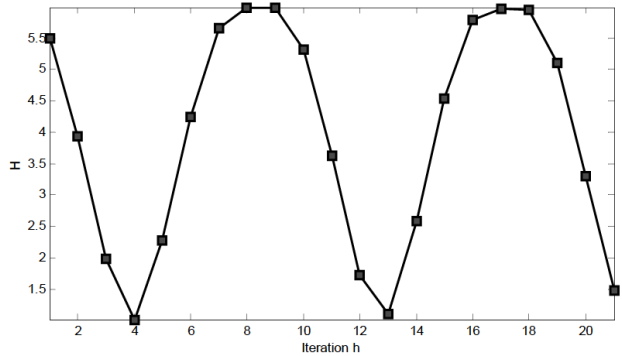

Figure 21: Shannon entropy analysis of Grover's QSA dynamics with five inputs.

Figure 21 shows that for five inputs of Grover's QSA an optimal number of iterations, according to minimum of the Shannon entropy criteria for successful result, is exactly four. After that probability of correct answer will decrease and algorithm may fail to produce correct answer. Note that theoretical estimation for 5 inputs gives $\frac{\pi}{4}\sqrt{2^5} = 4.44$ iterations.

Simulation results of fast Grover QSA are summarized in Table 2.

Numbers of iterations for fast algorithm were estimated according to termination condition as minimum of Shannon entropy of quantum state vector. The following approaches were used in simulation:

*Approach 1:* Quantum operators are applied as matrices; elements of quantum operator matrices are calculated dynamically according to Eqs. (5), (6), and (7). Classical Hardware limit of this approach is around 20 qubits, caused by exponential temporal complexity.

*Approach 2:* Quantum operators are replaced with classical gates. Product operations are removed from simulation according to [8]. State vector of probability amplitudes is stored in compressed form (only different probability amplitudes are allocated in memory). With second approach it is possible to perform classical efficient simulation of Grover's QSA with arbitrary large number of inputs (50 qubits and more).

With allocation of the state vector in computer memory, this approach permits to simulation 26 qubits on PC with 1GB of RAM. Figure 22 shows memory required for Grover algorithm simulation, when whole state vector is allocated in memory.

| $n$ | Number of iterations $h$ | Temporal complexity, seconds | |
|---|---|---|---|
| | | Approach 1 (one iteration) | Approach 2 ($h$ iterations) |
| 10 | 25 | 0.28 | ~0 |
| 12 | 50 | 5.44 | ~0 |
| 14 | 100 | 99.42 | ~0 |
| 15 | 142 | 489.05 | ~0 |
| 16 | 201 | 2060.63 | ~0 |
| 20 | 804 | - | ~0 |
| 30 | 25.375 | - | 0.016 |
| 40 | 853.549 | - | 4.263 |
| 50 | 26.353.589 | - | 12.425 |

Table 2: Temporal complexity of Grover's QSA simulation on 1.2GHz computer with two CPUs.

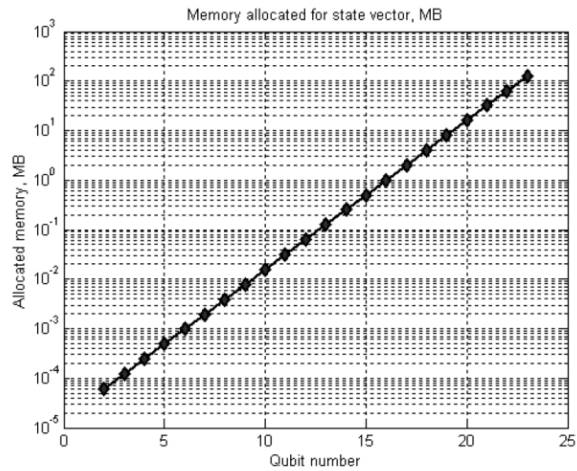

Figure 22: Spatial complexity of Grover QA simulation.

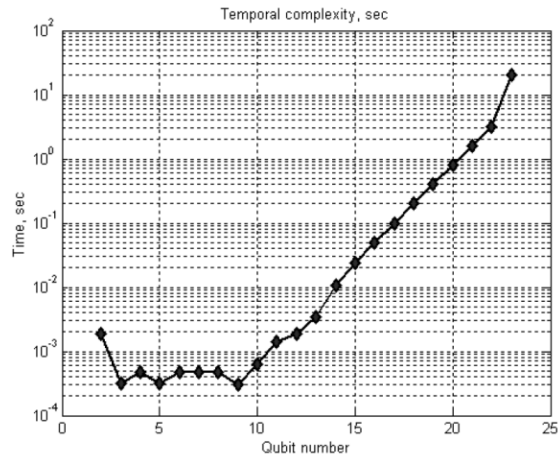

Figure 23: Temporal complexity of Grover's QSA.

Adding one qubit require double of the computer memory needed for simulation of Grover's QSA in case when state vector is allocated completely in memory.

Temporal complexity of Grover's QSA is presented in Fig. 23.

In this case state vector is allocated in memory, and quantum operators are replaced with classical gates according to [8]. Fastest case is when we compress state vector and replace quantum operator matrices with corresponding classical gates according with [8]. In this case we obtain speedup according to Approach 2.

## 9 Fast QSA models: Structure and acceleration method of quantum algorithm simulation

The analysis of the quantum operator matrices that was carried out in the previous sections forms the basis for specifying the structural patterns giving the background for the algorithmic approach to QA modeling on classical computers. The allocation in the computer memory of only a fixed set of tabulated (pre-defined) constant values *instead* of allocation of huge matrices (even in sparse form) provides computational efficiency. Various elements of the quantum operator matrix can be obtained by application of an appropriate algorithm based on the structural patterns and particular properties of the equations that define the matrix elements. Each representation algorithm uses a set of table values for calculating the matrix elements. The calculation of the tables of the predefined values can be done as part of the algorithm's initialization.

### 9.1 Algorithmic representation of the Grover's QA

Figures 24 (a - c) are flowcharts showing realization of such an approach for simulation of superposition (Fig. 24 (a)), entanglement (Fig. 24 (b)) and interference (Fig. 24 (c)) operators in Grover's QSA.

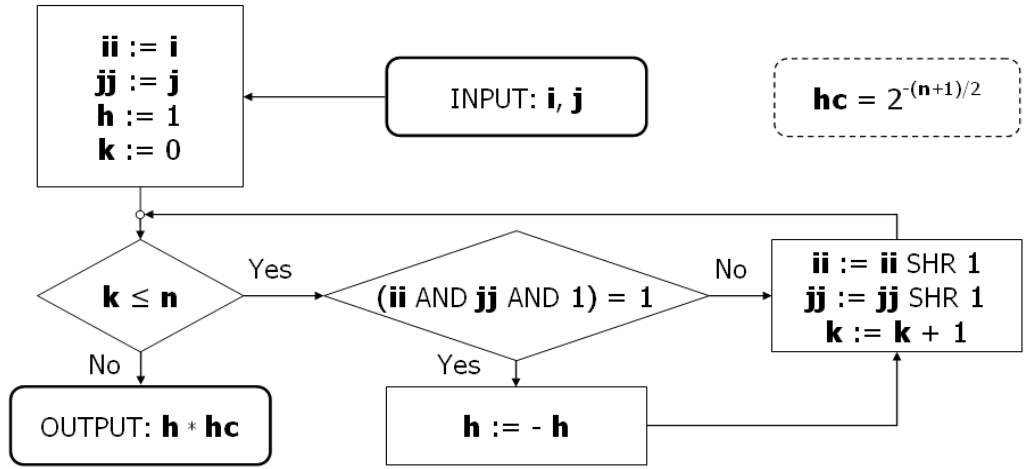

Figure 24 (a): Superposition operator representation algorithm for Grover's QSA.

Here $n$ is a number of qubit, $i$ and $j$ are the indexes of a requested element, $hc = 2^{-(n+1)/2}$, $dc1 = 2^{1-n} - 1$ and $dc2 = 2^{1-n}$ are the table values.

In Fig. 24 (a), the *i,j* values are specified and provided to an initialization block with loops control variables *ii*:= *i*, *jj*:= 0, and *k*:= 0 are initialized, and calculation variable *h*:= 1 is initialized. The process then proceeds to a decision block. In the decision block, if $k$ is less than or equal to $n$, then the process advances to another decision block; otherwise, the process advances to an output block, where the output $h*hc$ is computed (where $hc = 2^{-(n+1)/2}$).

In the decision block, if ($ii$ and $jj$ and 1) = 1, then the process advances to a block $h := -h$; otherwise, the process advances to another block and passes to the next iteration without probability amplitude inversion. Alternatively, the process sets $h := -h$ and proceeds to the next iteration. By setting $ii := ii$ SHR 1, $jj := jj$ SHR 1, and $k := k + 1$ (where SHR is a shift right operation), and then the process continues until all probability amplitudes are assigned.

In Fig. 24 (b), the inputs $i, j$ in an input block are initialized as $ii := i$ SHR 1, and $jj := $ SHR 1 and then are passed to the end test.

If the end test is not succeeding, means the inputs $i$ and $j$ are pointing to the marked elements, the process of the probability amplitude inversion of the marked states in this case is performed.

In Fig. 24 (c), interference operator of Grover's QSA can be substituted by a simple logic which outputs 0 *if* (($i$ XOR $j$) AND 1) = 1 then regarding nonzero elements, *if* $i = j$ then the process outputs $dc1$, otherwise the process outputs $dc2$, where $dc1 = 2^{1-n} - 1$ and $dc2 = 2^{1-n}$.

The superposition and entanglement operators for Deutsch-Jozsa's QA are the same with superposition and entanglement operators for Grover's QSA (Figs 24 (a), and 24 (b), respectively).

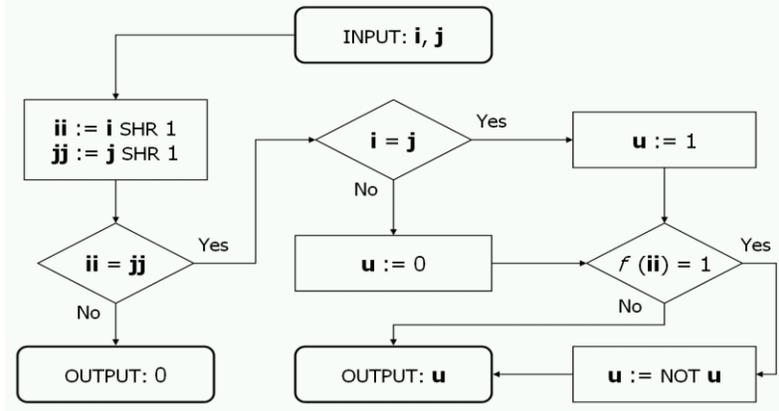

Figure 24 (b): Entanglement operator representation algorithm for Grover's QSA.

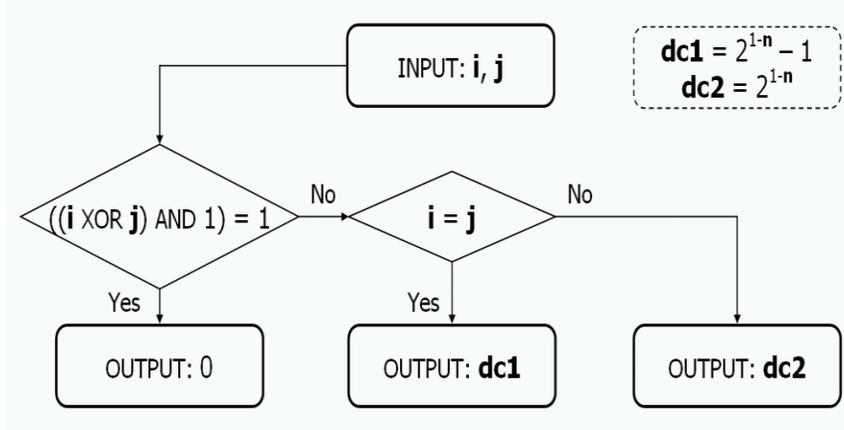

Figure 24 (c): Interference operator representation algorithm for Grover's QSA.

The time required for calculating the elements of an operator's matrix during a process of applying a quantum operator is generally small in comparison to the total time of performing a quantum step. Thus, the time burden created by exponentially increasing memory usage tends to be less, or at least similar to, the time burden created by computing matrix elements as needed. Moreover, since the algorithms used to compute the matrix elements tend to be based on fast bit-wise logic operations, the algorithms are amenable to hardware acceleration.

Table 3 shows comparisons of the traditional and as-needed matrix calculation when the memory used for the as-needed algorithm (Memory* denotes memory used for storing the quantum system state vector).

| Qubits | Standard | | Calculated Matrices | |
|---|---|---|---|---|
| | Memory, MB | Time, s | Memory* | Time, s |
| 1 | 1 | 0.03 | $\approx 0$ | $\approx 0$ |
| 8 | 18 | 3.4 | 0.008 | 0.0325 |
| 11 | 1048 | 1411 | 0.064 | 3.3 |
| 16 | -- | -- | 2 | 4573 |
| 24 | -- | -- | 512 | $3*10^8$ |
| 64 | -- | -- | -- | -- |

Table 3: Different approaches comparison: Standard (matrix based) and algorithmic based approach.

*The results shown in Table 3 are based on the results of testing the software realization of Grover QSA simulator on a personal computer with Intel Pentium III 1 GHz processor and 512 Mbytes of memory. Only one iteration of the Grover QSA was performed.

Table 3 shows that significant speed-up is achieved by using the algorithmic approach as compared with the prior art direct matrix approach. The use of algorithms for providing the matrix elements allows considerable optimization of the software, including the ability to optimize at the machine instructions level. However, as the number of qubits increases, there is an exponential increase in temporal complexity, which manifests itself as an increase in time required for matrix product calculations.

Use of the structural patterns in the quantum system state vector and use of a problem-oriented approach for each particular algorithm can be used to offset this increase in temporal complexity. By way of explanation, and not by way of limitation, the Grover algorithm is used below to explain the problem-oriented approach to simulating a QA on a classical computer.

## 9.2 Problem-oriented approach based on structural pattern of QA state vector

Let $n$ be the input number of qubits. In the Grover algorithm, half of all $2^{n+1}$ elements of a vector making up its even components always take values symmetrical to appropriate odd components and, therefore, need not be computed.

Odd $2^n$ elements can be classified into two categories:
- The set of $m$ elements corresponding to truth points of input function (or oracle); and
- The remaining $2^n - m$ elements.

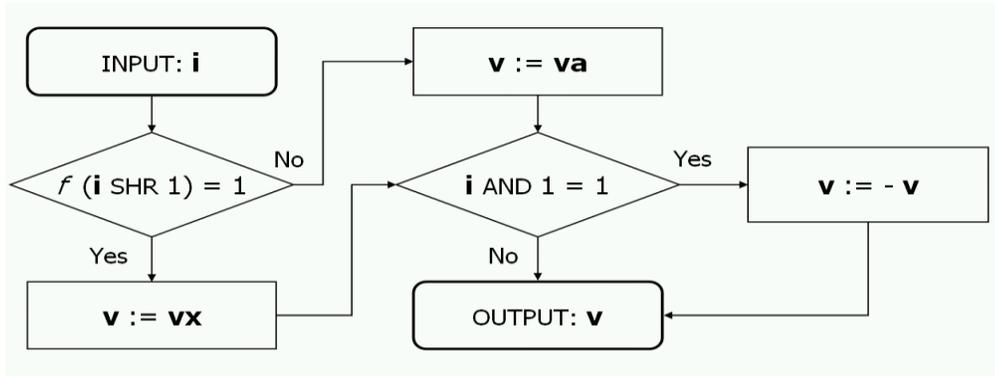

Figure 25: State vector representation algorithm for Grover' quantum search.

The values of elements of the same category are always equal.

As discussed above, the Grover QA only requires two variables for storing values of the elements. Its limitation in this sense depends only on a computer representation of the floating-point numbers used for the state vector probability amplitudes. For a double-precision software realization of the state vector representation algorithm, the upper reachable limit of q-bit number is approximately 1024.

Figure 25 shows a state vector representation algorithm for the Grover QA.

In Fig. 25, $i$ is an element index, $f$ is an input function, $vx$ and $va$ corresponds to the elements' category, and $v$ is a temporal variable. The number of variables used for representing the state variable is constant. A constant number of variables for state vector representation allow reconsideration of the traditional schema of quantum search simulation.

Classical gates are used not for the simulation of appropriate quantum operators with strict one-to-one correspondence but for the simulation of a quantum step that changes the system state. Matrix product operations are replaced by arithmetic operations with a fixed number of parameters irrespective of qubit number.

Figure 26 shows a generalized schema for efficient simulation of the Grover QA built upon three blocks, a superposition block $H$, a quantum step block UD and a termination block $T$.

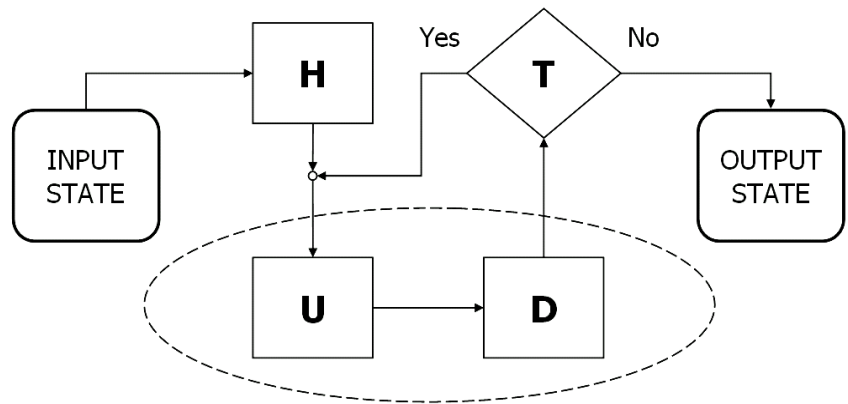

Figure 26: Generalized schema of simulation for Grover' QSA.

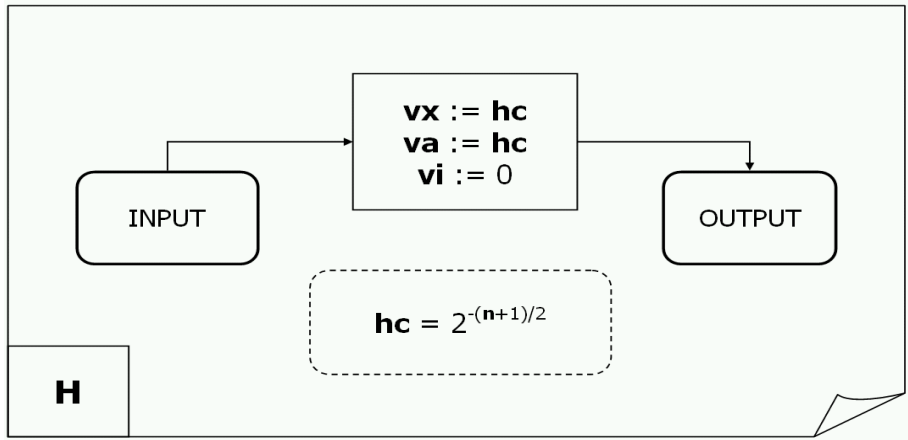

Figure 27: Superposition block for Grover's QSA.

Figure 26 also shows an input block and an output block. The $UD$ block includes a $U$ block and a $D$ block. The input state from the input block is provided to the superposition block. A superposition of states from the superposition block is provided to the $U$ block. An output from the $U$ block is provided to the $D$ block. An output from the $D$ block is provided to the termination block. If the termination block terminates the iterations, then the state is passed to the output block; otherwise, the state vector is returned to the U block for iteration.

As shown in Fig. 27, the superposition block $H$ for Grover QSA simulation changes the system state to the state obtained traditionally by using $n + 1$ times the tensor product of Walsh-Hadamard transformations. In the process shown in Fig. 24, *vx:= hc, va:= hc,* and *vi:= 0*, where $hc = 2^{-(n+1)/2}$ is a table value.

The quantum step block UD that emulates the entanglement and interference operators is shown on Figs 28 (a - c).

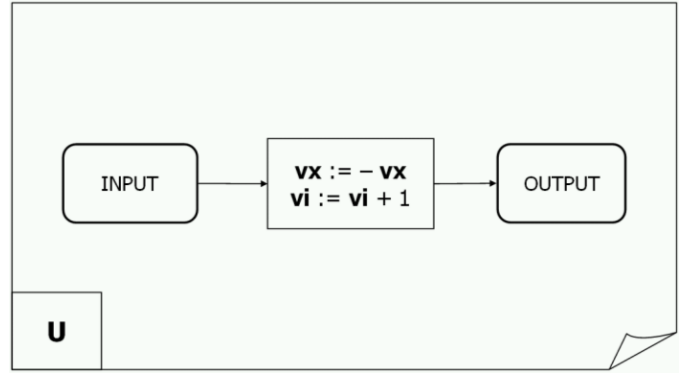

Figure 28 (a): Emulation of the entanglement operator application of Grover's QSA.

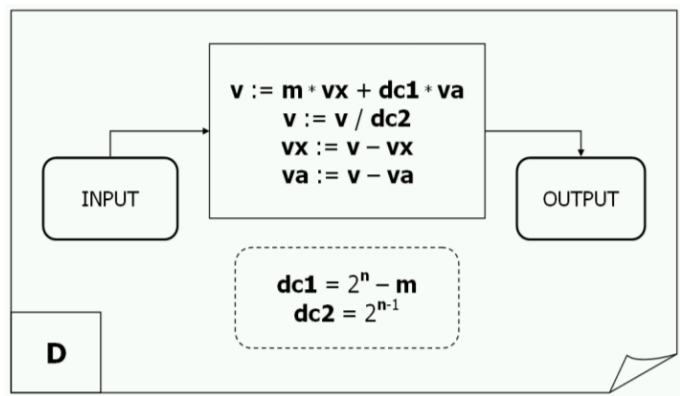

Figure 28 (b): Emulation of interference operator application of Grover's QSA.

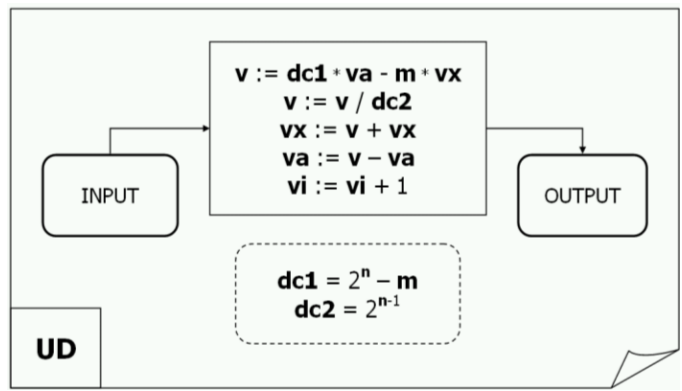

Figure 28 (c): Quantum step block for Grover' quantum search.

The *UD* block reduces the temporal complexity of the quantum algorithm simulation to linear dependence on the number of executed iterations.

The *UD* block uses recalculated table values $dc1 = 2^{1-n} - 1$ and $dc2 = 2^{1-n}$.

In the *U* block shown in Fig. 28 (a), $vx := -vx$ and $vi := vi + 1$.

In the *D* block shown in Fig. 28 (b), $v := m*vx + dc1*va$, $v := v/dc2$, $vx := v - vx$, and $va := v - va$ in the *UD* block shown in Fig. 28 (c), $v := dc1*va - m*vx$, $v := v/dc2$, $vx := v + vx$, $va := v - va$, and $vi := vi + 1$.

The termination block *T* is general for all QAs, independently of the operator matrix realization. Block *T* provides *intelligent termination condition* for the search process. Thus, the block *T* controls the number of iterations through the block *UD* by providing enough iteration to achieve a high probability of arriving at a correct answer to the search problem. The block *T* uses a rule based on observing the changing of the vector element values according to two classification categories. The *T* block during a number of iterations, watches for values of elements of the same category monotonically increase or decrease while values of elements of another category changed monotonically in reverse direction. If after some number of iterations, the direction is changed, it means that an extremum point corresponding to a state with maximum or minimum uncertainty is passed. The process can apply direct values of amplitudes instead of considering Shannon entropy value, thus, significantly reducing the required number of calculations for determining the minimum uncertainty state that guarantees the high probability of a correct answer.

The termination algorithm realized in the block T can be used one or more of five different termination models:

*Model* 1: Stop after a predefined number of iterations;
*Model* 2: Stop on the first local entropy minimum;
*Model* 3: Stop on the lowest entropy within a predefined number of iterations;
*Model* 4: Stop on a predefined level of acceptable entropy; and/or
*Model* 5: Stop on the acceptable level or lowest reachable entropy within the predefined number of iterations.

Note that models 1 - 3 do not require the calculation of an entropy value.

Figures 29 - 31 show the structure of the termination condition blocks *T*.

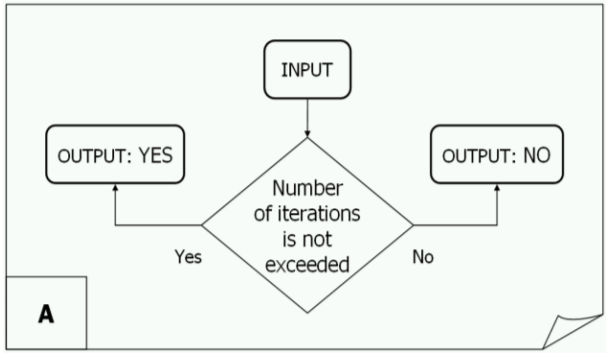

Figure 29: Termination block for method 1.

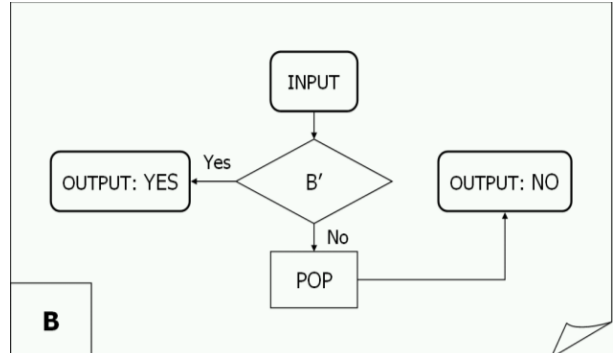

Figure 30: Component B for the termination block.

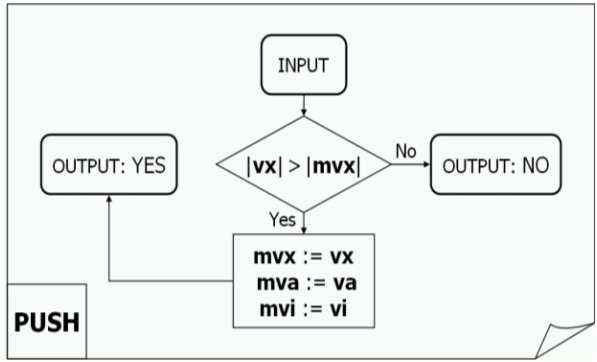

Figure 31 (a): Component PUSH for the termination block.

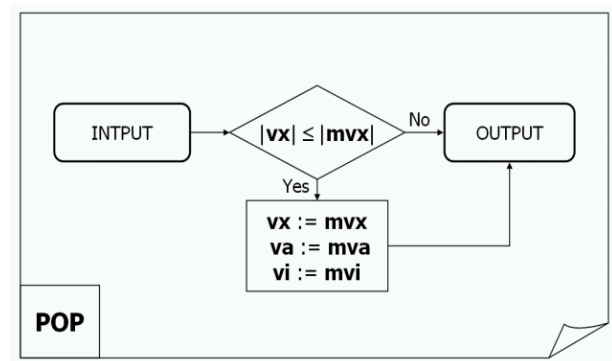

Figure 31 (b): Component POP for the termination block.

Since time efficiency is one of the major demands on such termination condition algorithm, each part of the termination algorithm is represented by a separate module, and before the termination algorithm starts, links are built between the modules in correspondence to the selected termination model by initializing the appropriate functions' calls.

Table 4 shows components for the termination condition block T for the various models. Flow charts of the termination condition building blocks are provided in Figs 29 - 31.

| Model | T | B' | C' |
|---|---|---|---|
| 1 | A | -- | -- |
| 2 | B | PUSH | -- |
| 3 | C | A | B |
| 4 | D | -- | -- |
| 5 | C | A | E |

Table 4: Termination block construction.

The entries A, B, PUSH, C, D, E, and PUSH in Table 3.6 correspond to the flowcharts in Figs 32 - 34 respectively.

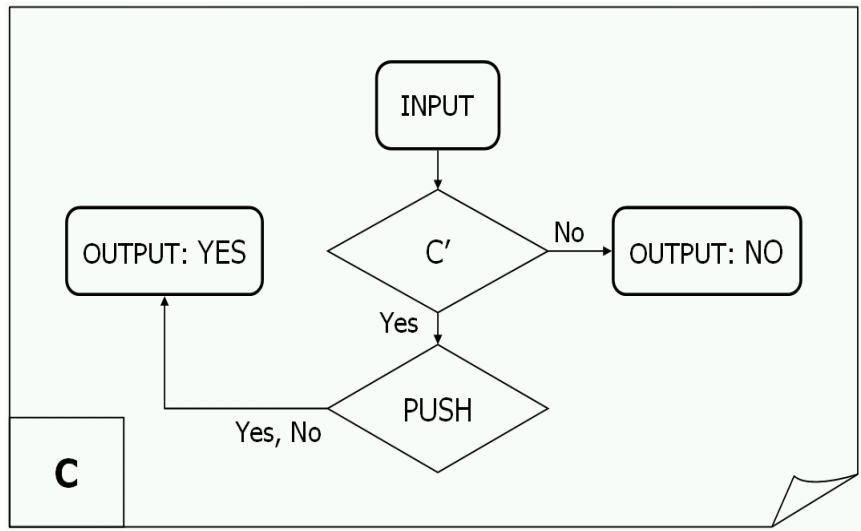

Figure 32: Component C for the termination block.

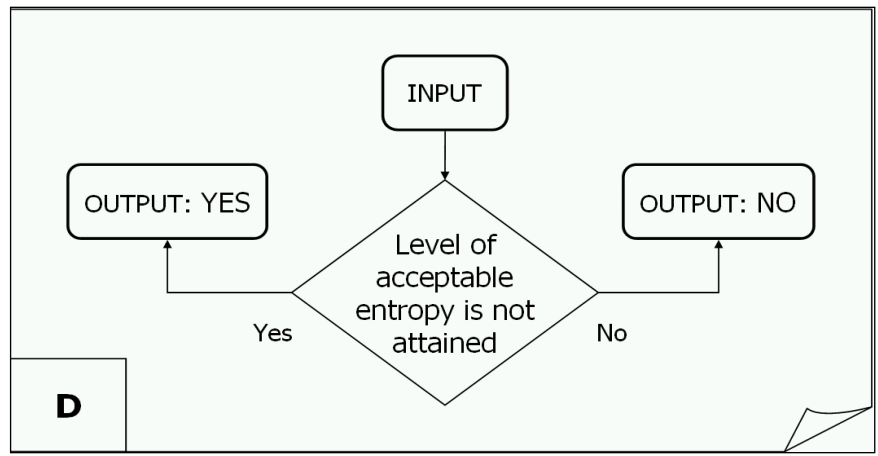

Figure 33: Component D for the termination block.

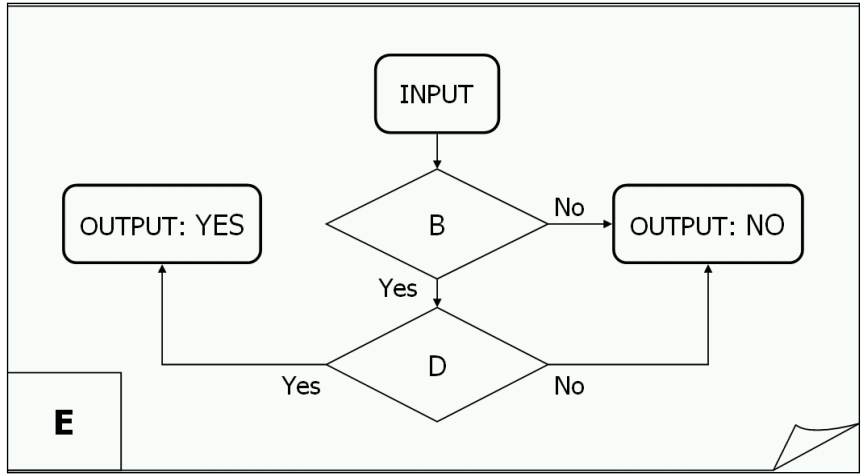

Figure 34: Component E for the termination block.

In *model* 1, only one test after each application of quantum step block *UD* is needed. This test is performed by block *A*. So, the initialization includes assuming *A* to be *T*, i.e., function calls to *T* are addressed to block *A*. Block *A* is shown in Fig. 29.

As shown in Fig. 29, the *A* block checks to see if the maximum number of iterations has been reached, if so, then the simulation is terminated, otherwise, the simulation continues.

In *model* 2, the simulation is stopped when the direction of modification of categories' values are changed. Model 2 uses the comparison of the current value of *vx* category with value *mvx* that represents this category value obtained in previous iteration:

(i) If *vx* is greater than *mvx*, its value is stored in *mvx*, the *vi* value is stored in *mvi*, and the termination block proceeding to the next quantum step;

(ii) If *vx* is less than *mvx*, it means that the *vx* maximum is passed and the process needs to set the current (final) value of *vx* := *mvx*, *vi* := *mvi*, and stop the iteration process. So, the process stores the maximum of *vx* in *mvx* and the appropriate iteration number *vi* in *mvi*. Here block *B*, shown in Fig. 30 is used as the main block of the termination process.

The block PUSH, shown in the Fig. 31 (a) is used for performing the comparison and for storing the *vx* value in *mvx* (*case a*). A POP block, shown in Fig. 31 (b) is used for restoring the *mvx* value (*case b*). In the PUSH block of Fig. 31 (a), *if |vx| > |mvx|*, then *mvx:= vx, mva:= va, mvi:= vi*, and the block returns true; otherwise, the block returns false.

In the POP block of Fig. 28 (b), *if |vx| <= |mvx|, then vx:= mvx, va:= mva*, and *vi:= mvi*.

The *model* 3 termination block checks to see that a predefined number of iterations do not exceed (using block *A* in Fig. 31):

- If the check is successful, then the termination block compares the current value of *vx* with *mvx*. If *mvx* is less than, it sets the value of *mvx* equal to *vx* and the value of *mvi* equal to *vi*. If *mvx* is less using the PUSH block, then perform the next quantum step;
- If the check operation fails, then (if needed) the final value of *vx* equal to *mvx*, *vi* equal to *mvi* (using the POP block) and the iterations are stopped.

The *model* 4, the termination block uses a single component block *D*, shown in Fig. 33.

The *D* block compares the current Shannon entropy value with a predefined acceptable level. If the current Shannon entropy is less than the acceptable level, then the iteration process is stopped; otherwise, the iterations continue.

The *model* 5 termination block uses the *A* block to check that a predefined number of iterations do not exceeded. If the maximum number is exceeded, then the iterations are stopped. Otherwise, the *D* block is then used to compare the current value of the Shannon entropy with the predefined acceptable level. If acceptable level is not attained, then the PUSH block is called and the iterations continue. If the last iteration was performed, the POP block is called to restore the *vx* category maximum and appropriate

*vi* number and the iterations are ended.

Figure 35 shows measurement of the final amplitudes in the output state to determine the success or failure of the search.

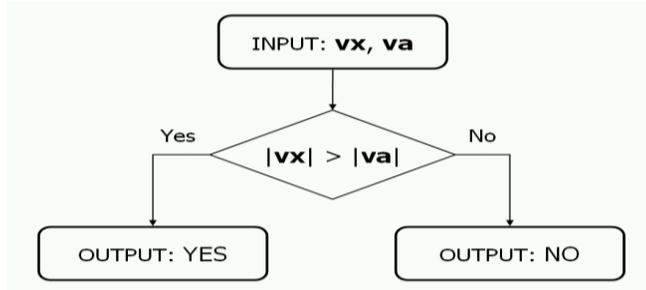

Figure 35: Final measurement emulation.

If $|vx| > |va|$, then the search was successful; otherwise, the search was not successful.

Table 5 lists results of testing the optimized version of Grover QSA simulator on personal computer with Pentium 4 processor at 2GHz.

| Qbits | Iterations | Time |
| --- | --- | --- |
| 32 | 51471 | 0.007 |
| 36 | 205887 | 0.018 |
| 40 | 823549 | 0.077 |
| 44 | 3294198 | 0.367 |
| 48 | 13176794 | 1.385 |
| 52 | 52707178 | 3.267 |
| 56 | 210828712 | 20.308 |
| 60 | 843314834 | 81.529 |
| 64 | 3373259064 | 328.274 |

Table 5: High probability answers for Grover QSA.

Using the above algorithm, a simulation of a 1000 qubit Grover QSA requires only 96 seconds for $10^8$ iterations (see below Fig. 36).

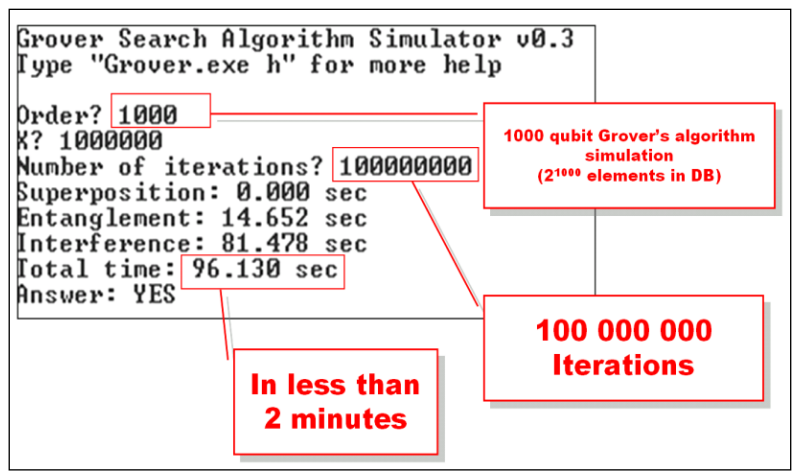

Figure 36. Simulation results of problem-oriented Grover QSA according to approach 4 with 1000 qubits.

The theoretical boundary of this approach is not the number of qubits, but the representation of the floating-point numbers.

The practical bound is limited by the front side bus frequency of the personal computer.

*Related works*. Presented approach was firstly suggested in [6-11] for efficient simulation of quantum algorithms on classical computers with minimum Shannon entropy measure of termination of searching processes [7] and differ from results in [12-20].

Acknowledgements. Fruitful discussion with P. Shor, Ch. Bennet, L. Levitin and V. Belavkin (especially in Capri, Italy, 2000 and MIT Boston, 2005) help us to formulate more clear presented results for quantum software engineering applications [21].

# Conclusions

- Design method of modular system for realization of Grover's Quantum Search Algorithm are presented.
- Design process of main quantum operators with algorithmic description for quantum algorithm gates simulation on classical computer is developed.
- Model representations of quantum operators in fast QAs introduced.
- Algorithmic based approach, when matrix elements are calculated on "demand" described.
- Problem-oriented approach demonstrated, where succeeded to run Grover's algorithm with up to 64 and more qubits with Shannon entropy calculation (up to 1024 without termination condition).

These results are the background for efficient simulation on classical computer the quantum soft computing algorithms, robust fuzzy control based on quantum genetic (evolutionary) algorithms and quantum fuzzy neural networks (that can realized as modified Grover's QSA), AI-problems as quantum game's gate simulation approaches and quantum learning, quantum associative memory, quantum optimization, etc.